\documentclass{emulateapj}
\usepackage{topcapt}
\shorttitle{Photometric and Spectroscopic Evolution of the IIP SN 2007it}
\shortauthors{Andrews et al.}
\usepackage{natbib}
\usepackage{rotating}

\begin{document}

\title{Photometric and Spectroscopic Evolution of the IIP SN 2007it to Day 944}

\author{J.E. Andrews\altaffilmark{1},  B.E.K. Sugerman\altaffilmark{2}, Geoffrey C. Clayton\altaffilmark{1}, J.S. Gallagher\altaffilmark{3}, M.J. Barlow\altaffilmark{4}, J. Clem\altaffilmark{1}, B. Ercolano\altaffilmark{5},  J. Fabbri \altaffilmark{4}, M. Meixner\altaffilmark{6},  M. Otsuka\altaffilmark{6}, D.L. Welch\altaffilmark{7}, and R. Wesson\altaffilmark{4}}
\altaffiltext{1}{Department of Physics and Astronomy, Louisiana State University, 202 Nicholson Hall, Baton Rouge, LA 70803; jandrews@phys.lsu.edu, jgallagher@phys.lsu.edu, gclayton@fenway.phys.lsu.edu, jclem@phys.lsu.edu}
\altaffiltext{2}{Department of Physics and Astronomy, Goucher College, 1021 Dulaney Valley Rd., Baltimore, MD 21204; ben.sugerman@goucher.edu}
\altaffiltext{3}{Department of Mathematics, Physics, and Computer Science, Raymond Walters Collge, 9555 Plainfield Rd., Blue Ash, OH 45236; gallagjl@ucmail.uc.edu}
\altaffiltext{4}{Department of Physics and Astronomy, University College London, Gower Street, London WC1E 6BT, UK; mjb@star.ucl.ac.uk, jfabbri@star.ucl.ac.uk, rwesson@star.ucl.ac.uk}
\altaffiltext{5}{Astrophysics Group, University of Exeter, Stocker Road, Exeter, EX4 4QL, UK; barbara@astro.ex.ac.uk}
\altaffiltext{6}{Space Telescope Science Institute, 3700 San Martin Drive, Baltimore, MD 21218; meixner@stsci.edu, otsuka@stsci.edu}
\altaffiltext{7}{Department of Physics and Astronomy, McMaster University, Hamilton, Ontario, L8S 4M1 Canada; welch@physics.mcmaster.ca}

\begin{abstract}
SN 2007it is a bright, Type IIP supernova which shows indications of both pre-existing and newly formed dust. The visible photometry shows a bright late-time luminosity, powered by the 0.09 M$_{\sun}$ of $^{56}$Ni present in the ejecta.  There is also a sudden drop in optical brightness after day 339, and a corresponding brightening in the IR due to new dust forming in the ejecta. CO and SiO emission, generally thought to be precursors to dust formation, may have been detected in the mid-IR photometry of SN 2007it. The optical spectra show stronger than average [O I] emission lines and weaker than average [Ca II] lines, which may indicate a 16 - 27 M$_{\sun}$ progenitor, on the higher end of expected Type IIP masses.  Multi-component [O I] lines are also seen in the optical spectra, most likely caused by an asymmetric blob or a torus of oxygen core material being ejected during the SN explosion. Interaction with circumstellar material prior to day 540 may have created a cool dense shell between the forward and reverse shocks where new dust is condensing. At late times there is also a flattening of the visible lightcurve as the ejecta luminosity fades and a surrounding light echo becomes visible.  Radiative transfer models of SN 2007it SEDs indicate that up to 10$^{-4}$ M$_{\sun}$ of new dust has formed in the ejecta, which is consistent with the amount of dust formed in other core collapse supernovae.
\end{abstract}


\section{Introduction}

Type II supernovae (SNe) are core collapse supernovae (CCSNe) whose spectra show the presence of hydrogen.  They are sub-divided based on light curve evolution, and Type IIP SNe show an extended plateau phase lasting $\sim$ 70-110 days due to radiative cooling of the shock-heated hydrogen envelope as well as energy released from hydrogen recombination \citep{1971Ap&SS..10...28G}. This phase is then followed by a tail of linear decline in luminosity, due to the radioactive decay of $^{56}$Co, with the luminosity being constrained by the amount of ejected  $^{56}$Ni \citep{1980NYASA.336..335W}. Type IIP SNe make up almost 60$\%$ of CCSNe \citep{2009MNRAS.395.1409S}. The progenitors of this subclass are thought to be red supergiants (RSG) with masses between 8-25 M$_{\sun}$ \citep{2004MNRAS.353...87E,2003ApJ...591..288H} that steadily lose their atmospheres with winds of 10-15 km s$^{-1}$. Recently, a comprehensive analysis by \citet{2009MNRAS.395.1409S} did not find any evidence for progenitors above 17 M$_{\sun}$, which may indicate a disconnect between theory and observation since theoretical models of CCSNe progenitors, such as those listed in \citet{2010MNRAS.tmp..284M}, have estimated progenitor masses of up to 29 M$_{\sun}$.  Therefore the search is ongoing for detection of more massive progenitors. 

The study of the role of CCSNe as major contributors to dust production in the universe has grown dramatically over the last few years. This is due to the discovery of high redshift galaxies containing massive amounts of dust  \citep{2003A&A...406L..55B}. The origin of this dust is still under much contention, whether it be from quick, explosive processes such as from CCSNe or from slow, quiescent processes such as from the atmospheres of asymptotic giant branch (AGB) stars. CCSNe, which can quickly return their material back into the surrounding ISM, are attractive potential producers of dust seen at high-z. Up to 1 M$_{\sun}$ of dust per SN would be needed to account for the dust seen at high-z according to the work of \citet{2001MNRAS.325..726T} and \citet{2003ApJ...598..785N}.  Numerous comprehensive studies of nearby SNe have been undertaken in order to estimate dust masses, but at the most only 10$^{-2}$ M$_{\sun}$ of newly formed dust has been detected \citep{2006Sci...313..196S}, with the average newly formed dust masses being 10$^{-3}$ - 10$^{-4}$ M$_{\sun}$ \cite[for example]{2003MNRAS.338..939E,2007ApJ...665..608M,2009ApJ...704..306K}. Low mass stars take much longer to evolve into the dust producing AGB phase, and may not have had enough time to evolve to become major dust producers in galaxies less than 1 Gyr old as they do in modern galaxies \citep{2007ApJ...662..927D,2010ApJ...712..942M}. Although, recent papers by \cite{2009MNRAS.397.1661V}, \cite{2009Sci...323..353S}, and \cite{2010arXiv1011.1303D} suggest AGB stars may be significant dust contributors as early as $\sim$150-500 Myr. \citet{2010ApJ...712..942M} find that in six submillimeter galaxies with z $>$ 4, the dust content of three can be explained by AGB stars, while the remaining three can only be explained by CCSNe.

Dust formation in the ejecta of CCSNe can be detected by certain observational signatures.  These include a decrease in the optical luminosity due to dust grains absorbing the visible light, while at the same time creating an excess in the IR as the grains re-emit the light at longer wavelengths.  Lastly, the spectral lines may appear to be asymmetric and blue-shifted as the dust grains obscure the receding (red) side of the ejecta more so than the approaching (blue) side.  All three of these signatures were seen for the first time in SN 1987A \citep{1989LNP...350..164L,1993ApJS...88..477W}, then in SN 2003gd \citep{2006Sci...313..196S,2007ApJ...665..608M}, and also SN 2004et \citep{2006MNRAS.372.1315S,2009ApJ...704..306K}. There have also been several CCSNe in the past few years that have shown one or two of these indicators of dust formation \citep[and references therein]{2009ASPC..414...43K}.  Although these signatures usually appear between 1 and 2 years after the explosion, there has recently been confirmation of dust forming much earlier (less than 200 days) in a few CCSNe with circumstellar interaction. Normal Type II RSG progenitors only lose about 10$^{-6}$ to 10$^{-5}$ M$_{\sun}$ yr$^{-1}$ \citep{2006ApJ...641.1029C}, which is too tenuous to show any circumstellar medium (CSM) interaction. The progenitors of Type IIn SNe on the other hand lose orders of magnitude more material, 10$^{-2}$ to 10$^{-1}$ M$_{\sun}$ yr$^{-1}$ \citep{2010arXiv1010.2689K}, which result in narrow ($\sim$ 100 km s$^{-1}$) emission lines in their spectra due to ionization of the pre-existing CSM which has been excited by the initial flash of the supernova.  For example SN 1998S showed dust formation signatures between days 140 - 268 \citep{2000ApJ...536..239L}, and SN 2005ip appears to have formed dust both in the CDS between day 75-150 and then again in the ejecta after day 750 \citep{2009ApJ...695.1334S,2009ApJ...691..650F}.  Although not classified as Type IIn, the Type Ib/c SN 2006jc also formed dust in the cool dense shell (CDS) created by the interaction of the ejecta and the CSM between 50 - 75 days post-explosion \citep{2008MNRAS.389..141M,2008ApJ...680..568S} and the Type IIP SN 2007od formed dust sometime between day 120 - 230 through the same mechanism \citep{2010ApJ...715..541A}.

In this paper, we follow the evolution of the Type IIP SN 2007it in the optical and IR from day 10 to day 944. This makes SN 2007it one of only a few Type IIP SNe with long-term and extensive spectral and photometric observations, along with SN 1990E, SN 1999em, SN 1999gi, SN 2002hh, SN 2003gd, SN 2004et, and SN 2005cs \citep[and references therein]{2010MNRAS.tmp..284M}. In Section 3 we discuss the visible lightcurve evolution, including a possible scattered optical light echo seen at late times.  In Section 4 we present the optical spectral evolution and describe the unusual spectral evolution of SN 2007it.  The IR lightcurve evolution is discussed in Section 5, followed by the radiative transfer modeling and SED fitting in Section 6. In all four sections we will discuss the evidence that dust has formed in the ejecta of SN 2007it.

\section{Observations and Analysis}

\objectname{SN 2007it} was discovered  in \objectname{NGC 5530} by R. Evans visually on 2007 September 13 with V $\sim$ 13.5 mag. \citep{2007CBET.1065....1E,2007IAUC.8874....3I}. Pre-discovery images taken with the All Sky Automated Survey (ASAS-3) constrain the explosion date between 2007 September 4-6 \citep{2007IAUC.8875....1P}, and for the purposes of this paper we are assuming an explosion date of 2007 September 5 (JD 2454348). It was confirmed spectroscopically to be a Type II SN by the Carnegie Supernova Project on 2007 September 15 \citep{2007CBET.1068....1C}. Using \citet{2008ApJ...676..184T}, which uses distances set by the 2001 HST Cepheid Key Project observations, we are adopting a distance of 11.7 Mpc throughout this paper.  However it should be noted that a previous study done by \citet{1988ngc..book.....T} had suggested a distance of 16.9 Mpc. We must also note here that no pre-explosion images exist of the host galaxy of SN 2007it.

We have obtained visible spectroscopy and photometry as well as mid-IR photometry of SN 2007it spanning days 107-944. Lists of these observations are presented in Tables 1 and 2.  Optical imaging was obtained in Johnson-Cousins BVI  with the SMARTS consortium 1.3m telescope at Cerro Tololo Inter-American Observatory (CTIO), Chile.  All images were pipeline reduced, shifted, and stacked. Imaging and spectra were also obtained with GMOS/Gemini South (GS-2008A-Q-24, GS-2008B-Q-45, GS-2009A-Q-49, GS-2010A-DD-3). The g$^{\prime}$r$^{\prime}$i$^{\prime}$ images were reduced and stacked using the IRAF \textit{gemini} package.  The instrumental g$^{\prime}$r$^{\prime}$i$^{\prime}$  magnitudes were transformed to standard Johnson-Cousins VRI \citep{2007ApJ...669..525W,1996AJ....111.1748F}.  For each night the transformation involved a least-squares fit with a floating zero point. 

One epoch of photometry was obtained with the WFPC2/PC1 camera on HST in the F450W, F606W, and F814W filters. These images were delivered pipeline reduced, but undrizzled, and stacking and cosmic ray removal was accomplished using the Pyraf task \textit{multidrizzle}.  Transformation into the standard Johnson-Cousins $BVI$ was done using methods outlined by \citet{1995PASP..107.1065H}.  Late time images were obtained with the Wide Field Camera (WFC) on HST/ACS using the F435W, F606W, and F814W filters. These images were pipeline reduced, including drizzling and cosmic ray removal, and transformations to the Johnson-Cousins $BVI$ system were accomplished using methods outlined by \citet{2005PASP..117.1049S}. 

A $BVRI$ photometric sequence of tertiary standard stars (shown in Figure 1) was derived for the SN 2007it field, using the same method as \cite{2010ApJ...715..541A}. The $BVRI$ magnitudes for these standards are located in Table 3.  The $BVRI$ lightcurves of SN 2007it are shown in Figure 2 and the absolute V magnitude is shown in comparison with other similar Type II SNe in Figure 3. Early time photometry of SN 2007it from days 9-20 (shown in Figures 2 and 3) was obtained from the Carnegie Supernova Project.   The discrepancies between the SMARTS and Gemini I magnitudes are likely due to the bandpass of the Cousins I filter covering part of the Ca II IR-triplet around 8500\AA\ that is not covered by the Gemini i$^{\prime}$ filter.  This would cause the SMARTS photometry to appear brighter due to the added Ca II flux. Uncertainties for the Gemini photometry were calculated by adding in quadrature the transformation uncertainty quoted in \citet{2007ApJ...669..525W}, photon statistics, and the zero point deviation of the standard stars for each epoch.
The HST uncertainties consist of the transformation uncertainty from \citet{1995PASP..107.1065H} or \citet{2005PASP..117.1049S} and photon statistics added in quadrature. 

For each GMOS/Gemini South epoch, three spectra of 900s (before day 540) or 1800s (day 540 and after) were obtained in longslit mode using grating B600 and a slit width of 0$\farcs$75. Central wavelengths of 5950, 5970, and 5990 \AA\ were chosen to prevent important spectral features from falling on chip gaps. A 2x2 binning in the low gain setting was used. Spectra were reduced using the IRAF \textit{gemini} package.  The sky subtraction regions were determined by visual inspection to prevent contamination from material not associated with the SN, and the spectra were extracted using 15 rows centered on the SN. The spectra from each individual night were averaged and have been corrected for the radial velocity of  \objectname{NGC 5530} (1196 km s$^{-1}$). They are presented in Figures 4 and 5.

 In order to estimate the reddening towards SN 2007it, we compared the day 10 spectrum with other Type IIPs of a similar epoch with known $E(B-V)$ values. The day 10 spectrum was most similar to SN 2007od and SN 1999em.  Given the quoted foreground reddening of  $A_{v}$ = 0.39 from NED for SN 2007it, and the total reddening by 0.39 $\pm$ 0.04 of SN 2007od  \citep{2010ApJ...715..541A} and 0.31 $\pm$ 0.16 for SN 1999em \citep[and references therein]{2010MNRAS.tmp..284M} we have estimated that  the reddening of the host galaxy is relatively low, and adopt $A_{v}$ = 0.39 for the purpose of this paper. We also first attempted to use the narrow Na I D absorption seen in the day 10 spectrum of SN 2007it to determine the foreground extinction. The equivalent width (EW) of the blended Na I D lines was 1.6 $\pm$ 0.1\AA, which corresponded to an A$_{v}\sim$1.27 $\pm$ 0.1 of the host galaxy \citep{1990A&A...237...79B}. This value does not agree with the day 10 spectrum which indicated that the SN is not reddened by a high amount.  Discrepancies between reddening measured from Na I EW and actual values are not unique. For example, \citet{2009ApJ...693..207B} found that for an Na I EW of $\sim$ 0.2 \AA\, $E(B-V)$ had values ranging between 0.2 and 1.5, and that one could not solely rely on EW values. Therefore we do not use the Na I reddening estimate in this paper.

The Spitzer IRAC (3.6, 4.5, 5.8, and 8.0$\mu$m) images were mosaicked and resampled using standard MOPEX procedures to improve photometric quality. PSF photometry was performed using the position specific PRF images.  Table 5 contains the measured mid-IR fluxes obtained from Spitzer. Figure 6 presents the SED of SN 2007it at four epochs, Spitzer (day 351)/Gemini (day 339), Spitzer (day 561)/Gemini (day 566), Spitzer (day 718)/HST (day 696) and Spitzer (day 944)/Gemini (day 916). Statistical uncertainties presented in the plot represent 1$\sigma$ errors.   The data have been corrected for foreground extinction.

\section{Lightcurve Evolution}

\subsection{Day 9 - Day 339}
 
The photometric evolution of SN 2007it at early times seems to be consistent with other Type IIP supernova lightcurves, as can be seen in Figure 3. The plateau phase appears to begin near day 20, and lasts until around the time that the SN becomes visible again on day 107 when it is in midst of its the post-plateau drop in brightness. Therefore the plateau phase lasted $\sim$ 80 days.  Given the distance to SN 2007it and the V magnitudes on day 20 and day 107,  the average absolute magnitude of SN 2007it on day $\sim$50 is estimated to be M$_{V}$=-16.7 (L = 1.6 x 10$^{42}$ erg s$^{-1}$), once corrections for reddening have been made. Comparing these values with other Type IIP SNe \citep[and references therein]{2010MNRAS.tmp..284M} indicates that SN 2007it is one of the more luminous, well studied Type IIP SNe to date and is quite similar to SN 2004et (see Figure 3). 

After the plateau phase, the drop to the radioactive tail is roughly 2 mag. Type IIP SN normally experience a decrease in luminosity of $\sim$1.5-3 mag based upon the amount of $^{56}$Ni present \citep{1986ARA&A..24..205W,2009ApJ...703.2205K}. For comparison, SN 2007it shows roughly the same plateau duration and drop as SN 2004et (Figure 3), which may suggest a similar $^{56}$Ni mass. This is discussed in more detail below. The decline in luminosity then follows the $^{56}$Co decay closely (see Figure 2), until the SN is lost behind the sun after day 339.

\subsection{Day 509 - Day 922}
Once SN 2007it was observed again on day 509, we found that its brightness had fallen below the expected $^{56}$Co decay by $\sim$ 0.5 mag in R. The R band contains the H$\alpha$ and [O I] $\lambda\lambda$ 6300,6364 \AA\ emission features, which are responsible for most of the flux at late times.  The most likely cause for this decreased decline in brightness is new dust formation between days 339 and 509.  This is the time period during which dust typically condenses in the ejecta of SNe.  After this drop, the $^{56}$Co decay was maintained, indicating that little or no further dust formation occurred after day 509. Other dust indicators such as the simultaneous increase in the IR luminosity and red-wing attenuation of the [O I] lines, discussed below, indicate that this is indeed the case.  

Also of note is another deviation from the $^{56}$Co decay curve seen as a flattening of the lightcurve after day 700.  This was also seen in SN 2007od \citep{2010ApJ...715..541A}, SN 2003gd \citep{2005ApJ...632L..17S}, and SN 2002hh \citep{2007ApJ...669..525W}, and is likely a light echo caused by dust reflecting light from the flash of the SN explosion, keeping the luminosity brighter than expected at late times. A comprehensive explanation of scattered light echoes is presented in \citet{2003AJ....126.1939S}.  Scattered light echoes are expected to become major contributors to the emission of SNe around 8-9 mag below maximum \citep{2005MNRAS.357.1161P}, which is when we see the flattening of the SN 2007it lightcurve. Analysis of this echo is beyond the scope of this paper, and will be presented in a forthcoming paper. Subtraction of images of SN 2007it taken with HST/ACS on day 922 and three epochs of WFC3 data (to presented in another paper) between days 857 and 1009 revealed a residual in the southwest corner roughly 4 pixels from the center of the SN corresponding to an echo from ISM material $\sim$21 pc away from the SN (Andrews et al. 2011, in preparation).  Comparison of photometry from early times to the late-time flattening, show the late-time color is much bluer, with  B-V dropping from $\sim$1.5 on day 107, to $\sim$ 0.3 on day 922. This will happen when the CSM or ISM dust is preferentially scattering the blue light as was seen in SN 2006gy on day 810 \citep{2010AJ....139.2218M}. The resolved light echo, along with the blue late-time colors, indicate that the flattening of the light curve after day 696 is likely due to a light echo and not CSM interaction at late times.

\section{Spectral Evolution}
 Figure 4 shows the optical spectral evolution of SN 2007it from day 10 to day 922. The early time spectra show broad H$\alpha$ and H$\beta$ emission and strong Na I D absorption. The H$\alpha$ evolution is seen in Figure 5, and shows a blueshift for at least the first 200 days. Blueshifted H$\alpha$ emission has been seen in other Type II SNe such as SN 1988A \citep{1993MNRAS.265..471T}, SN 1998A \citep{2005MNRAS.360..950P}, and SN 1999em \citep{2003MNRAS.338..939E}. \citet{1988SvAL...14..334C} and later \citet{1990sjws.conf..149J} proposed that this was caused by the reflection of photons by the photosphere.  For 1988A and 1999em, the blueshift only lasted $\sim$80 days, whereas SN 2007it and SN 1998A seem to have a blueshift persisting well past 150 days. By day 200, when the SN is well into the nebular phase,  strong [O I] $\lambda\lambda$ 6300,6363 \AA\  emission is seen (shown in Figure 5), as well as [Ca II] $\lambda\lambda$7291,7324 and [Fe II] $\lambda$7155 \AA\  emission. 
 
 Once we recover SN 2007it at day 509, the [O I] line strengths rival that of H$\alpha$, while the [Ca II] lines are at most, half the strength of H$\alpha$. This is unusual for Type IIP SNe, since most exhibit stronger [Ca II] than [O I]. The only other example of  a Type II SN with similar late-time line strengths as SN 2007it, was the Type IIL SN 1970G \citep{1976SvA....20..666P,1988Ap&SS.146..375C}. Unfortunately no optical spectra were taken after $\sim$ day 350 of SN 1970G to allow further comparisons of the spectral evolution. There is also an emission feature emerging after day 200 at 7380 \AA\, which is separate from the [Ca II]/[O II] blended lines, and could be a [Ni II] $\lambda$7380 \AA\ line which was also identified in the Type IIn SN 2006gy \citep{2009ApJ...697..747K}, Type Ic SN 2006aj \citep{2007ApJ...658L...5M}, and a few type Ia SNe \citep{2010ApJ...708.1703M}. This line may be present in other SNe but hidden by strong [Ca II] emission \citep{2007ApJ...658L...5M}.  

 Of special note are the multi-component peaks of the [O I] $\lambda \lambda$6300,6364 \AA\ lines that are visible in our day 154 spectra and persist to day 622 (Figure 6).  These emission line profiles are similar to the multiple peaks seen in the H$\alpha$ emission of SNe 1998S \citep{2000ApJ...536..239L} and 2007od \citep{2010ApJ...715..541A}. Both  the 6300 and 6364 \AA\ lines have a blue-shifted component occurring at -1300 km s$^{-1}$ in relation to the component seen at the rest velocity of the galaxy.  There seems to be no corresponding red peaks at +1300 km s$^{-1}$. Because CSM interaction is normally only seen in H and He lines, this suggests another origin for the multiple peaks.
 Similar asymmetries in the [O I] lines have been seen in numerous H-poor CCSNe (Type Ib, Ic, and IIb), for example SNe 2005aj, 2006T, and 2008ax, but seems to be rare in normal Type IIP \citep{2008ApJ...687L...9M,2009MNRAS.397..677T,2010ApJ...709.1343M}. These asymmetries can be seen  approximately 2 months post explosion in H-poor CCSNe, but due to the lack of spectral observations between days 10 and 154 in SN 2007it, we cannot pinpoint the exact date before day 154 when the [O I] lines and the corresponding asymmetries appeared.  \citet{2010ApJ...709.1343M} suggest that the the absence of a red component combined with the presence of blue peak the same distance away from both the 6300 \AA\ and 6364 \AA\ lines could be explained by emission from an asymmetric blob of O material ejected on the forward side of the SN during the explosion.  They do caution that a torus of O material is less likely but cannot be ruled out, and that the red-shifted components could exist but are hidden either by the SN orientation or  obscuring dust.   The existence of fast moving knots seen in oxygen-rich SN remnants such as Puppis A and Cas A may substantiate the asymmetric blob scenario. According to \citet{1985ApJ...299..981W}, it is likely that these knots are intact fragments of the core of the progenitor star. These remnant oxygen knots have radial velocities consistent with the blue [O I] components of SN 2007it.
   
  Signatures of CSM interaction can also be seen in the H$\alpha$ and [O I] lines (Figures 5 and 6). The CSM interaction is manifested as a narrower feature (FWHM $\sim$ 1000 km s$^{-1}$) at the rest velocity rising above the broader (FWHM $\sim$4000 km s$^{-1}$) ejecta base. Although this feature does not become apparent until day 509 in the H$\alpha$ spectra, it seems to be present since day 154 in the [O I] emission.  We then surmise that the intermediate peak was present in both lines since day 154, but the overwhelming ejecta emission in H$\alpha$ kept it hidden until day 509. From the day 154 spectra and onward in [O I] there also seems to be an attenuation of the red-wing  of the central peak, especially in 6300 \AA\ which becomes even more apparent after day 509 when the asymmetric blob emission has faded. There are two possibilities for these asymmetries, attenuation from pre-existing dust or Bochum events like that seen in H$\alpha$ of SNe 1987A and 1999em.  For a full discussion on these events see  \citet{2003MNRAS.338..939E}. The shape of the [O I] central peaks are similar to modeled profiles presented in \citet{2003MNRAS.338..939E}, where the peak has been shifted toward the blue side due to the presence of dust in the ejecta. As we will show below, IR data from day 351 suggests the presence of an IR echo from flash-heated pre-existing dust.  This pre-existing dust is likely the cause of the attenuation seen in the [O I] lines. We do not see signatures of new dust forming from asymmetries in the spectral lines, although we must point out that it is possible that if the new dust is concentrated in a ring or torus around the SN, newly formed dust could be hidden by the viewing angle.  If the newly formed dust has formed in the cool dense shell between the forward and reverse shocks formed in the CSM interaction, the ejecta emission would be located interior to the newly formed dust, which would allow equal attenuation of the blue and red emission.

\subsection{Mass estimates}
We estimate a $^{56}$Ni mass for SN 2007it of  M$_{Ni}$= 0.09$^{+.01}_{-.02}$ M$_{\sun}$ from the methods of \citet{2003ApJ...582..905H}.  The bolometric luminosity of the radioactive tail (in erg s$^{-1}$) is calculated as 
\begin{equation}
log_{10}L_t= \frac{-[V_t-A(V)+BC] + 5log_{10}D-8.14}{2.5},
\end{equation}
with A$_{V}$ = 0.39 mag, and a bolometric correction of BC = 0.26. The nickel mass is then calculated at various times during this phase, as
\begin{equation} 
M_{Ni} = (7.866 x 10^{-44})L_texp[\frac{\frac{t_t-t_0}{1+z}-6.1}{111.26}]M_{\sun}.
\end{equation}
Here 6.1 and 111.26 days are the half-lives of $^{56}$Ni and $^{56}$Co, respectively, and $t_{t}$-$t_{0}$ is the age of the SN. This calculated Ni mass is similar, if slightly higher, than other Type IIP SNe.  SN 2004et has an estimated M$_{Ni}$ mass of 0.056 M$_{\sun}$\citep{2010MNRAS.tmp..284M}, SNe 1999em, 2003gd, and 2004dj each have $\sim$0.02 M$_{\sun}$ \citep{2003MNRAS.338..939E, 2005MNRAS.359..906H, 2006MNRAS.369.1780V}, and the low-luminosity SN 2005cs contains 3 x 10$^{-3}$M$_{\sun}$ of $^{56}$Ni \citep{2009MNRAS.394.2266P}.

We have estimated the mass of oxygen two separate ways using the strength of the [O I] doublet in the nebular phase before the onset of dust formation. Using a temperature range of 3500-4000 K, and a flux (F$_{[OI]}$) of 9.7 x 10$^{-14}$ erg s$^{-1}$ cm$^{-2}$ for the day 286 spectrum (obtained from the Carnegie Supernova Project) we estimate M$_{O}$= 0.2-0.9 M$_{\sun}$ using
\begin{equation}
M_{[OI]} = 10^8F_{[OI]}D^2e^{2.28/T_4}
\end{equation}
which is Equation 1 in  \cite{1986ApJ...310L..35U}. As a second method we implemented the relationship given in  \cite{2003MNRAS.338..939E},
\begin{equation}
L_{[OI]} = \eta \frac{M_O}{M_{exc}}L_{Co}
\end{equation}
under the assumption that SN 2007it had the same $\eta$ (the efficiency of energy transfer from O mass to [O I] doublet luminosity) and excited mass as SN 1987A, and that the L$_{Co}$, the luminosity of the Co emission, was directly related to the mass of $^{56}$Ni. Once again we used an oxygen flux of 9.7 x 10$^{-14}$ erg s$^{-1}$ cm$^{-2}$.  This calculation gives us an oxygen mass of 0.96 that of SN 1987A, or between 1.15-1.44 M$_{\sun}$.  These values, although not consistent with each other, are very similar to the values calculated for SN 2004et in \citet{2010MNRAS.tmp..284M}.

Comparing the relationship between $^{56}$Ni and progenitor mass in other SNe (for example, Table 5 in \citet{2007MNRAS.381..280M}), it is likely that SN 2007it had a main sequence mass $\geq$ 20 M$_{\sun}$.  Recently, \cite{2010arXiv1006.2268D} have suggested that using the velocity of the H$\alpha$ line at $\sim$ day 15 and the velocity of the [O I]  $\lambda\lambda$ 6300,6363 \AA\ lines at $\sim$ 300 days could be an indicator of progenitor ZAMS mass. Using the values of 10960 km s$^{-1}$ and 2000 km s$^{-1}$, respectively, and the estimated mass of oxygen from above suggests a progenitor mass between 16-20 M$_{\sun}$. \citet{1987ApJ...322L..15F,1989ApJ...343..323F}  have also suggested that the ratio between [Ca II]/[O I] lines in the nebular phase can be an indicator of progenitor mass, with smaller ratios belonging to higher mass progenitors. This ratio is 0.7 for SN 2007it between days 209-351, which is quite small for a Type IIP SN.  \citet{2004A&A...426..963E} showed that SN with ratios similar to SN 2007it are normally of Type Ib/c.  SN 1987A had a ratio of $\sim$ 3 from day 250-450 and  SN 2005cs meanwhile showed a ratio of 4.2 $\pm$0.6 on day 334 \citep{2009MNRAS.394.2266P}, which is 40$\%$ larger than SN 1987A, and over 6 times greater than SN 2007it.  This may indicate that the progenitor of SN 2007it may have been more massive than either SN 2005cs (7-13 M$_{\sun}$) and SN 1987A (14-20 M$_{\sun}$), with roughly a mass of 20-27 M$_{\sun}$. Even though each method yields different dust masses, and uncertainties in distance and reddening may add more ambiguity to these estimates, they all suggest that SN 2007it had a progenitor mass of $\sim$ 20$^{+7}_{-4}$M$_{\sun}$. Unfortunately, no images of the progenitor star exist, although as was shown in the case for SN 2004et, masses from progenitor images do not necessarily always agree with other estimates from explosion parameters \citep[and references therin]{2010MNRAS.tmp.1772C, 2010MNRAS.tmp..284M}.

\section{ Mid-IR Evolution}
SN 2007it was monitored in the mid-IR for four epochs spanning days 351- 944 (see Table 5).  As can be seen in Figure 7, the first epoch on day 351 shows an indication of an increased flux in the 4.5 $\mu$m band with respect to the other channels.  This elevated emission in this band photometry was also seen around the same time period in SNe 2005af \citep{2006ApJ...651L.117K}, 2004dj \citep{2005ApJ...628L.123K} and 2004et \citep{2009ApJ...704..306K}, and has been attributed to CO fundamental band emission.  Less prominent is the possible extra emission at 8.0 $\mu$m which may be due to SiO emission. SiO emission was discovered in 1987A \citep{1991MNRAS.252P..39R} and also in SN 2004et \citep{2009ApJ...704..306K}, and 2005af \citep{2006ApJ...651L.117K}.  CO and SiO are thought to be precursors to dust formation \citep{2003ApJ...598..785N} since they likely provide the molecular foundation from which the dust grains form.  Therefore the presence of these molecules is important in understanding the whole story of dust formation in SNe.  

Day 351 shows the presence of an IR excess, well before the estimated time of new dust formation as indicated by visible photometry.  A likely cause for this elevated flux is an IR echo from flash-heated pre-existing CSM dust. Comparison with the models of SN 1980K presented in \citet{1983ApJ...274..175D}, which also had a peak luminosity of $\sim$  10$^{9}$ L$_{\sun}$, indicates a dust free cavity extending to  3 x 10$^{17}$ cm with a shell of material extending up to or past  6.3 x 10$^{17}$ cm.  This corresponds to a plateau of luminosity lasting from day $\sim$ 200 up to day 643, depending on the model. We do not have any IR observations prior to day 351, so we cannot definitively suggest the early behavior, but it is likely that an IR excess was present early on and we are seeing the continued plateau of this heated dust on day 351, similar to that seen in SN 1980K. If, though, we use the optical spectra on day 339 before the CSM interaction is apparent in H$\alpha$ and day 540 when the intermediate peak is visible along with the ejecta velocity on day 10 of 10 960 km s$^{-1}$ \citep{2007CBET.1068....1C}, we estimate that the initial SN flash has cleared a dust-free cavity between 3.2 and 5.3 x 10$^{16}$ cm from the center.  This agrees with estimates for SN 2005ip, which has the same L$_{peak} \sim$ 10$^{9}$ L$_{\sun}$, and a vaporization radius of $\sim$ 10$^{16}$ cm \citep{2010arXiv1005.4682F}.  Assuming a cavity was cleared out to 4 x 10$^{16}$ cm, the optical and UV light would reach this boundary at around 15 days and the time, $t$ for the IR plateau to last would only be $2R/c$, or $\sim$ 31 days. Although we cannot definitively explain the discrepancy between spectral constraints and SN 1980K model comparisons, it is possible that where \citet{1983ApJ...274..175D,1985ApJ...297..719D} considers a single, very thin shell of CSM material, SN 2007it may have a much more extended shell or multiple shells, creating a much later and/or longer plateau.

The IR fluxes on day 561 are very different from day 351, most importantly the fluxes in the 3.6, 5.8 and 8.0 $\mu$m channels have doubled in intensity. This may indicate that new dust grains have formed. This is likely since this is the same time period in which we see a dimming in the visible lightcurve (between days 339 and 540). The combination of the elevated mid-IR flux at the same time as the corresponding dimming in the visible can be explained if dust grains are absorbing the high energy photons and re-emitting them in the IR.  This has been seen in other dust producing CCSNe such as SN 1987A \citep{1991AJ....102.1118S} and SN 2003gd \citep{2006Sci...313..196S}.   Modified blackbody fits to this epoch indicate a blackbody luminosity of 4.5 x 10$^{38}$ erg s$^{-1}$, twice the blackbody luminosity at day 351. This is consistent with the optical drop of 0.5 mag between the two epochs which causes the early time luminosity to be 1.58 times brighter, or almost twice the brightness of the luminosity on day 540.  Within  the uncertainties, the drop in optical luminosity is equal to the increase in IR luminosity.

Days 718 and 922 were both observed by Spitzer in warm mode, so only the 3.6 and 4.5 $\mu$m bands were available.  Both epochs show significant IR luminosity, likely due to the newly formed ejecta dust. In the day 922 data, the 4.5 $\mu$m flux appears to be stronger than expected, and in fact the ratio between 3.6 and 4.5 is much smaller than other epochs,  only 0.27 as opposed to 0.6 on day 718.  If we assume the ratio to be constant in the day 922 epoch, we would expect a flux of $\sim$ 66 $\mu$Jy, but instead see a flux of 143 $\mu$Jy, indicating that there is possibly an excess in the 4.5 $\mu$m channel.  This is also seen by Sugerman et al. (2010, In preparation) for SN 2004et. It is likely we are still seeing CO emission that was visible on day 351, which was hidden by the increased emission from the new dust on day 561, and became visible again in the later epochs as the grains cooled.

\section{ Radiative Transfer Modeling}

In order to quantify the amount and composition of the dust in SN 2007it we used our 3D Monte Carlo radiative transfer code MOCASSIN \citep[and references therein]{2005MNRAS.362.1038E}. We have chosen to model three different geometries in which the dust is distributed either uniformly within a spherical shell surrounding the SN, with the addition of clumps scattered throughout the shell, or in a torus at some inclination around the SN. These will be known as  ``smooth'', ``clumpy'', and ``torus'' models, respectively. Relying on previous modeling done by \citet{2006Sci...313..196S}, \cite{2007ApJ...665..608M}, and \cite{2009ApJ...704..306K} we have used a standard MRN grain size distribution of $a^{-3.5}$ between 0.005 and 0.05 $\mu$m \citep{1977ApJ...217..425M} for all three scenarios. Tables 6 and 7, and Figure 7 show the model results.  

As we assumed for SN 2007od \citep{2010ApJ...715..541A}, in these models the dust and luminosity for the source was located between inner radius R$_{in}$ and outer radius R$_{out}$ of a spherically expanding shell, with the luminosity being proportional to the density at each location. For day 351 we have chosen to concentrate the luminosity in a point source, rather than spread throughout the shell, as it is likely this emission is the result of flash-heated pre-existing CSM. This does create a better fit to the IR data points. The smooth model assumes the density of dust in the shell was inversely proportional to the square of the radius. For the clumpy model the photons originate in the inhomogeneous interclump medium, where the clumps are considered to be optically thick and spherical.  For the torus models, densities are specified for the inner and outer walls, with the dust distribution falling off linearly between the two radii \citep{2007MNRAS.375..753E}. 

For each model we used a combination of amorphous carbon (AC) and silicate grains, using the optical constants of \citet{1988ioch.rept...22H} and \cite{1984ApJ...285...89D}, respectively.  For all epochs the best fit occurred with a carbon rich combination of 75$\%$ AC and 25$\%$ silicates, as was also the case in SN 2007od  \citep{2010ApJ...715..541A}. For day 351 no attempt was made to fit the 4.5 $\mu$m point, since it was likely due mainly to CO emission, the same is true for day 922 when CO may also be responsible for the elevated 4.5 $\mu$m flux. 

Luminosity, ejecta temperature, inner and outer radii, and dust masses were all inputs for each model.  Initial estimates for each parameter were accomplished using black body fits to the optical and IR data. This yielded dust temperatures (T$_{d}$) of roughly 500 K for the first epoch, and 700K for day 561, 590 K for day 718 and 480 K for day 944. These temperatures are consistent with typical warm dust temperatures.  For our first epoch, we used an R$_{in}$ of 3.2 x 10$^{16}$ cm, which roughly corresponds to the evaporation radius of the initial flash of the SN.   All other epochs used an R$_{in}$ of 7 x 10$^{15}$ cm which created the best fits.  We also kept the temperature of the ejecta (T$_{ej}$) at 5700K for all epochs, since this is a reasonable ejecta temperature and does an adequate job fitting the first epoch of optical points.  The combination of high H$\alpha$ flux and [O I] emission lines in the R band relative to the continuum, and at later times additional flux due to the light echo has made it unlikely that the visible photometry is consistent with any reasonable blackbody temperature.  Keeping those two values constant, we then varied the luminosity, outer radii, and dust masses to get the most accurate fits. For all epochs we found that the clumpy distribution predicted on average slightly higher dust masses than the smooth and torus models. 

We estimate a dust mass for our first epoch of 5 x 10$^{-4} M_{\sun}$ for the smooth model, 7.3 x 10$^{-4} M_{\sun}$ for clumpy, and 1.6 x 10$^{-4} M_{\sun}$ for the torus.  The remaining epochs yielded smooth dust masses of 7.0 x 10$^{-5} M_{\sun}$ for day 561, 8.0 x 10$^{-5} M_{\sun}$ for day 718, and 4.6 x 10$^{-5} M_{\sun}$ for day 922. The larger dust mass from the first epoch might be explained by the flash heating of pre-exisiting CSM dust in an area much further away and roughly 200K cooler than following epochs. At this first epoch R$_{in}$ = 3.2 x 10$^{16}$ cm, which is larger than the R$_{out}$ of the remaining epochs. This pre-existing dust will cool, as described in Section 5 above, and fade after its initial heating by the SN flash so that by day 561 the IR SED is dominated by newly formed ejecta dust.  After day 561, R$_{in}$ = 7 x 10$^{15}$ and R$_{out}$ = 2 x 10$^{16}$, which likely places the dust at this time in the CDS interior to the point of CSM interaction.  These results also indicate that the amount of new dust created is smaller than the amount in the CSM.

On day 351,the MOCASSIN models estimate that $\tau_{v}$ = 0.08.  In order to independently estimate the optical depth, we used the methods of \citet{2009ApJ...691..650F} which uses total peak energy and IR energy($\tau \sim \frac{E_{IR}}{E_{IR}+E}$).  The peak luminosity of SN 2007it was $\sim$ 10$^{9}$ L$_{\sun}$, which would indicate E $\sim$ 10$^{49}$ erg using the estimates of SN 2005ip contained in \citet{2009ApJ...691..650F}.  The IR energy for the first 351 days is then 1 x 10$^{48}$ erg, using an L$_{bb}$  of 8.8 x 10$^{6}$ L$_{\sun}$.   This yields a $\tau$ = 0.09, consistent with our MOCASSIN fits.  Between day 351 and 561 we also find a significant increase in $\tau _{v}$, which jumps from 0.08 to 1.27 as the new dust forms. This is largely due to the order of magnitude decrease in R$_{out}$ between the two epochs, requiring similar amounts of dust in a much smaller volume. Although the optical light curve suggests only an A$_{V}$ = 0.5, considerations of geometry, such as the newly formed dust existing in a torus around the SN for which we are seeing inclined at an angle away from edge-on, could explain the higher amount estimated from MOCCASIN which would provide the total optical depth summed over the whole system, not just along our line of sight.  This scenario could also explain the lack of increased attenuation in the [O I] and H$\alpha$ spectral lines between days 339 and 540 when we see the other dust formation signatures, mainly the simultaneous increase in optical extinction and IR luminosity.  Therefore the total mass of new dust formed in SN 2007it appears to be $\sim$ 1.0 x 10$^{-4} M_{\sun}$, which although consistent with dust masses of other CCSNe, is still considerably smaller than the amount needed to account for the dust seen at high-z \citep{2003MNRAS.343..427M}.  

\section{Summary}
This paper presents the results of a comprehensive study of SN 2007it for almost three years post-explosion.  SN 2007it is a Type IIP supernova, most similar to SN 2004et, that shows some interesting characteristics and has produced $\sim$ 1.0 x 10$^{-4} M_{\sun}$ of dust in its ejecta, likely in the cool dense shell created from CSM interaction.  High late-time luminosities indicate that M$_{Ni}$= 0.09 M$_{\sun}$ was synthesized in the explosion, greater than seen in other SNe such as 1987A and 2004et. The [Ca II]/[O I] ratios post plateau are also much smaller than normally seen in a Type IIP, 0.70 prior to dust formation and 0.50 after.  These two factors, a higher than normal Ni mass and a small [Ca II]/[O I], along with the expansion velocity of H$\alpha$ at early times and of [O I]  at later times suggest that SN 2007it may have had a progenitor mass $\geq$ 20M$_{\sun}$, which is quite large for a Type IIP SN.  The late-time lightcurve and color evolution also suggests that a light echo exists around SN 2007it which will be discussed in detail in an upcoming paper. Estimates of oxygen mass suggest that 1.1-1.4 M$_{\sun}$ may be present, which is consistent with the Type II SNe 2004et and 1987A.  The oxygen lines also show strange behavior not previously seen in a Type II SNe.  The post-plateau spectra show a blueshifted component at -1300 km s$^{-1}$ to both the 6300 and 6364 \AA\ lines, with no corresponding peak at +1300 km s$^{-1}$. This is likely due to emission from a torus of oxygen rich material surrounding the SN or an asymmetric blob of O material ejected on the forward side of the SN during the explosion.

Modeling done on SN 2007it using our MOCCASIN radiative transfer code indicates that there was pre-existing dust surrounding the SN prior to explosion, and that up to $\sim$ 1.0 x 10$^{-4} M_{\sun}$ formed in the ejecta after day 351. Up to 1 M$_{\sun}$ of dust would need to be produced per SN to account for the dust amounts seen at high-z according to the work of \citet{2001MNRAS.325..726T} and \citet{2003ApJ...598..785N}. Recently, however, \citet{2010ApJ...713....1C} have suggested that these models may overestimate these masses, which would be consistent with the values seen in nearby SNe. Even if the amount of dust produced in SN 2007it may not be large enough to account for the dust budget of early galaxies \citep{1989ApJ...344..325K,2001MNRAS.325..726T}, it is on par with other SNe such as 2007od \citep{2010ApJ...715..541A}, 2004et \cite{2009ApJ...704..306K}, and SN 2006jc \cite{2008MNRAS.389..141M}.  Although the sample is still small, this ever-growing body of evidence seems to indicate that low-mass CCSNe such as the ones seen today are likely not major contributors to dust in the early universe.  In fact, \citet{2009Sci...323..353S} and \citet{2009MNRAS.397.1661V} have both suggested that AGB stars can be more significant contributors to dust formation in early galaxies than once thought and \citet{2010MNRAS.403..474W} found that SN 2008S was likely a massive AGB star, not a supernova at all, which produced a significant amount of dust. SNe may then only play a small part in the dust formation in early galaxies, but more long-term and comprehensive observations of SNe are needed to definitely answer the role of SNe as dust produces in high-z galaxies.\\
\\
We would like to thank the anonymous referee for the valuable suggestions that have improved this paper.  We would also like to thank Nidia Morell and the rest of the Carnegie Supernova Project for allowing us access to the early time spectoscopic data, as well as Parvis Ghavamian and Nathan Smith for helpful comments. This work has been supported by NSF grant AST-0707691 and NASA GSRP grant  NNX08AV36H. This work was supported by Spitzer Space Telescope RSA 1415602 and RSA 1346842,  issued by JPL/Caltech. A portion of this data was obtained at the Gemini Observatory, which is operated by the Association of Universities for Research in Astronomy (AURA) under a cooperative agreement with the NSF on behalf of the Gemini partnership. The standard data acquisition has been supported by NSF grants AST-0503871 and AST-0803158 to A. U. Landolt.

\begin{table*}[htbp]
  \centering
  \topcaption{Photometry Summary} 
  \begin{tabular}{cccccc} 
Day & JD & Telescope & Instrument  & Exposures & Exposure Time \textit{(s)}\\
\hline
107&2454455& SMARTS 1.3m &ANDICAM& 5& 20\\
113&2454461& SMARTS 1.3m &ANDICAM&  5& 20\\
139&2454487& SMARTS 1.3m &ANDICAM&  5& 20\\
157&2454505& SMARTS 1.3m &ANDICAM&  5& 20\\
188&2454536& SMARTS 1.3m &ANDICAM&  5& 20\\
209&2454557& Gemini South& GMOS Imaging&3&20\\
223&2454571& SMARTS 1.3m &ANDICAM&  5& 20\\
242&2454590& Gemini South& GMOS Imaging&3&20\\
252&2454600& SMARTS 1.3m &ANDICAM&  3& 100\\
276&2454624& SMARTS 1.3m &ANDICAM&  3& 100\\
279&2454627& Gemini South& GMOS Imaging&3&20\\
300&2454648& SMARTS 1.3m &ANDICAM&  3& 100\\
302&2454650& Gemini South& GMOS Imaging&3&20\\
314&2454662& SMARTS 1.3m &ANDICAM&  3& 100\\
339&2454687& Gemini South& GMOS Imaging&3&20\\
351&2454699&Spitzer&IRAC&12&100\\
540&2454888&Gemini South&GMOS Imaging&1&60\\
561&2454909&Spitzer&IRAC&12&100\\
566&2454914&Gemini South&GMOS Imaging&2&60\\
589&2454937&Gemini South&GMOS Imaging&2&60\\
613&2454961&HST&WFPC2&4&412\\
622&2454970&Gemini South&GMOS Imaging&2&60\\
696&2455044&HST&ACS/WFC&4&412\\
718&2455066&Spitzer&IRAC&12&100\\
916&2455264&Gemini South&GMOS Imaging&2&60\\
922&2455270&HST&ACS/WFC&4&424\\
  \hline
  \hline
  \end{tabular}
  \label{tab:booktabs}
\end{table*}

\begin{table*}[htbp]
  \centering
  \topcaption{Spectroscopy Summary} 
  \begin{tabular}{cccccc} 
Day & JD & Telescope & Instrument  & Exposures & Exposure Time \textit{(s)}\\
\hline
154&2454502& Gemini South& GMOS Spectra&3&900\\
178&2454526& Gemini South& GMOS Spectra&3&900\\
209&2454557& Gemini South& GMOS Spectra&3&900\\
242&2454590& Gemini South& GMOS Spectra&3&900\\
279&2454627& Gemini South& GMOS Spectra&3&900\\
302&2454650& Gemini South& GMOS Spectra&3&900\\
339&2454687& Gemini South& GMOS Spectra&3&900\\
540&2454888&Gemini South&GMOS Spectra&3&1800\\
566&2454914&Gemini South&GMOS Spectra&3&1800\\
589&2454937&Gemini South&GMOS Spectra&3&1800\\
622&2454970&Gemini South&GMOS Spectra&3&1800\\
916&2455264&Gemini South&GMOS Spectra&3&1800\\
 \hline
  \hline
  \end{tabular}
  \label{tab:booktabs}
\end{table*}

\begin{table*}[h!]
\caption{Tertiary BVRI Standards for NGC 5530}
\centering
\begin{tabular}{ccccc}
\hline
\hline
Star&B&V&R&I\\
\hline
A&17.743 $\pm$ 0.012&17.060 $\pm$ 0.001&16.655 $\pm$ 0.012&16.272 $\pm$ 0.017\\
B&19.115 $\pm$ 0.039&17.667 $\pm$ 0.011&16.696 $\pm$ 0.015&15.648 $\pm$ 0.013\\
C&19.235 $\pm$ 0.040&18.099 $\pm$ 0.015&17.485 $\pm$ 0.014&16.945 $\pm$ 0.025\\
D&18.057 $\pm$ 0.014&17.254 $\pm$ 0.021&16.772 $\pm$ 0.033&16.209 $\pm$ 0.088\\
E&18.652 $\pm$ 0.029&17.219 $\pm$ 0.016&16.240 $\pm$ 0.007&15.154 $\pm$ 0.011\\
F&17.672 $\pm$ 0.012&16.818 $\pm$ 0.009&16.318 $\pm$ 0.008&15.812 $\pm$ 0.014\\
G&18.281 $\pm$ 0.017&17.587 $\pm$ 0.033&17.205 $\pm$ 0.020&16.778 $\pm$ 0.028\\
\hline
\end{tabular}
\centering
\end{table*}

\begin{table*}[htbp]
  \centering
  \topcaption{Optical Photometry of SN 2007it} 
  \begin{tabular}{ccccc} 
  \hline
Day & B&V & R & I\\
\hline
107& 16.16 $\pm$ 0.10&14.70 $\pm$ 0.04 &- & 13.55 $\pm$ 0.04\\
113&16.83 $\pm$ 0.11&15.16 $\pm$ 0.08&-&13.90 $\pm$ 0.06\\
139&17.43 $\pm$ 0.04&15.88 $\pm$ 0.03&-&14.60 $\pm$ 0.03\\
157&17.59 $\pm$ 0.04&16.1 $\pm$ 0.03&-&14.77 $\pm$ 0.03\\
188&17.77 $\pm$ 0.05&16.37$\pm$ 0.02&- &15.04 $\pm$ 0.03\\
209&-&16.73 $\pm$ 0.04&15.71 $\pm$ 0.03 &15.67 $\pm$ 0.03\\
223&17.97 $\pm$ 0.03&16.73 $\pm$ 0.03&-&15.37 $\pm$ 0.03\\
242&-&17.01 $\pm$ 0.04&16.02 $\pm$ 0.02 &15.99 $\pm$ 0.03\\
252&18.52 $\pm$ 0.05&17.00 $\pm$ 0.03&-&15.75$\pm$ 0.09\\
276&18.35 $\pm$ 0.04&17.23 $\pm$ 0.03&- &15.90 $\pm$ 0.03\\
279&-&17.34 $\pm$ 0.04&16.42 $\pm$ 0.02 &16.33 $\pm$ 0.03\\
300&18.52 $\pm$ 0.05&17.44 $\pm$ 0.03&- &16.15 $\pm$ 0.03\\
302&-&17.54 $\pm$ 0.04&16.66 $\pm$ 0.03 &16.54 $\pm$ 0.02\\
314&18.64 $\pm$ 0.05&17.62 $\pm$ 0.05&- &16.36 $\pm$ 0.04\\
339&-&17.88 $\pm$ 0.04&17.09 $\pm$ 0.04 &16.89 $\pm$ 0.05\\
540&-&20.11 $\pm$ 0.04&19.60 $\pm$ 0.03&19.43$\pm$ 0.03\\
566&-&20.27 $\pm$ 0.04&19.78 $\pm$ 0.04 &19.62 $\pm$ 0.05\\
589&-&20.55 $\pm$ 0.04&20.05 $\pm$ 0.02&19.94$\pm$ 0.03\\
613&21.41 $\pm$ 0.16&20.59 $\pm$ 0.11 &- &19.93 $\pm$ 0.09\\
622&-&20.7 $\pm$ 0.04&20.32 $\pm$ 0.03 &20.10 $\pm$  0.03\\
696&21.81 $\pm$ 0.12&21.38 $\pm$ 0.05&-&20.95 $\pm$ 0.05 \\
916&-&22.02 $\pm$ 0.03&21.73 $\pm$ 0.03&21.81 $\pm$ 0.04 \\
922&22.16 $\pm$ 0.18&21.82 $\pm$ 0.10&-&20.95 $\pm$ 0.14 \\

  \hline
  \hline
  \end{tabular}
  \label{tab:booktabs}
\end{table*}

\begin{table*}[htbp]
\centering
\topcaption {Spitzer Photometry of SN 2007it}
\begin{tabular}{ccccc}
\hline
Day&3.6$\mu$m ($\mu$Jy)&4.5$\mu$m ($\mu$Jy)&5.8$\mu$m ($\mu$Jy)&8.0$\mu$m ($\mu$Jy)\\
\hline
351&257.28 $\pm$ 6.64 &625.75 $\pm$ 12.10 &317.40 $\pm$ 13.16&363.41 $\pm$ 23.28 \\
561&528.69 $\pm$ 11.69&600.90 $\pm$ 11.07  &660.69 $\pm$ 17.65  &628.06 $\pm$ 23.14 \\
718&213.21 $\pm$ 6.11 &354.16 $\pm$ 7.15 &- &-\\
944&39.12 $\pm$ 2.30 &142.96 $\pm$ 3.28 &- &-\\
\hline
\hline
\end{tabular}

\end{table*}

\begin{table}[h]
\centering
\topcaption {Monte Carlo Radiative Transfer Models}
\begin{tabular}{lccccccccc}
\hline
\multicolumn{2}{c}{} &
\multicolumn{6}{r}{Smooth} &
\multicolumn{1}{c}{} &
\multicolumn{1}{c}{Clumpy} \\
Epoch & AC/Si & T$_{ej}$ (K) & R$_{in}$ (cm) & R$_{out}$(cm) & L$_{tot.}$ (L$_{\odot}$) & $\tau$$_{v}$ & M$_{d}$ (M$_{\odot}$) & & M$_{d}$ (M$_{\odot}$) \\
\hline
351 d & 0.75/0.25 & 5700 &3.2e16 &9.0e17  & 1.9e7 & 0.08 & 5.0e-4 & & 7.3e-4 \\
561 d & 0.75/0.25 & 5700 &7.0e15 & 2.0e16 & 4.6e6 & 1.29 & 7.0e-5 & & 1.0e-4 \\
718 d & 0.75/0.25 & 5700 & 7.0e15 & 2.0e16 & 2.4e6 & 1.4 & 8.0e-5 & &  1.3e-4\\
944 d & 0.75/0.25 & 5700 & 7.0e15 &2.0e16 & 6.9e5 & 0.98 & 4.6e-5 & &  1.3e-4\\
\hline
\hline
\end{tabular}
\end{table}

\begin{table}[h]
\centering
\topcaption {Monte Carlo Radiative Transfer Torus Models}
\begin{tabular}{lcccccccc}
\hline
Epoch & AC/Si & T$_{ej}$ (K) & R$_{in}$ (cm) & R$_{out}$(cm) & L$_{tot.}$ (L$_{\odot}$) & $\tau$$_{v}$ & M$_{d}$ (M$_{\odot}$) \\
\hline
351 d & 0.75/0.25 & 5700 &3.2e16 &2.5e17  & 1.9e7 & 0.08 & 1.6e-4  \\
561 d & 0.75/0.25 & 5700 & 7.0e15 & 2.4e16 & 4.6e6 & 1.27 & 7.9e-5  \\
718 d &0.75/0.25 & 5700 & 7.0e15 & 2.4e16 & 2.4e6 & 1.27& 7.9e-5\\
944 d & 0.75/0.25 & 5700 & 7.0e15 & 2.4e16 & 6.9e5 & 1.27 & 7.9e-5\\
\hline
\hline
\end{tabular}
\end{table}

\clearpage
\newpage

\begin{figure}[htp]
\centering
  \includegraphics [width= 3 in] {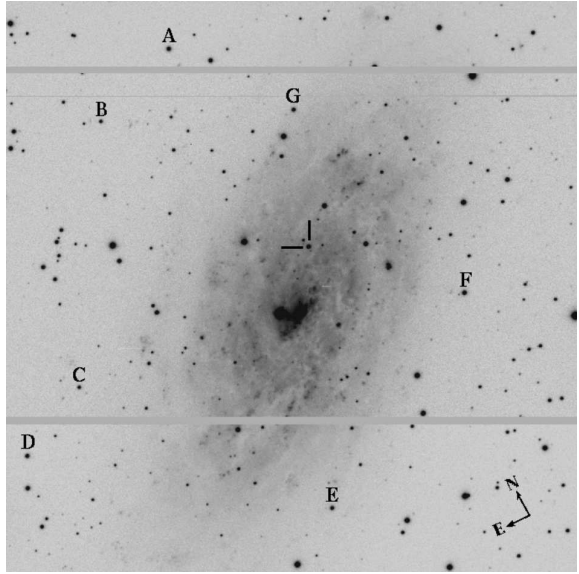}
  \caption{Image of NGC 5530 taken in the g$^{\prime}$ band with Gemini South on 2008 April 12. The tertiary standards listed in Table 3 are labeled alphabetically, and the SN is indicated at the center of the image. The image is 4.2' x 4.2'.}
  \end{figure} 
  
  \begin{figure}[htp]
\centering
  \includegraphics [width= 3 in,angle=90] {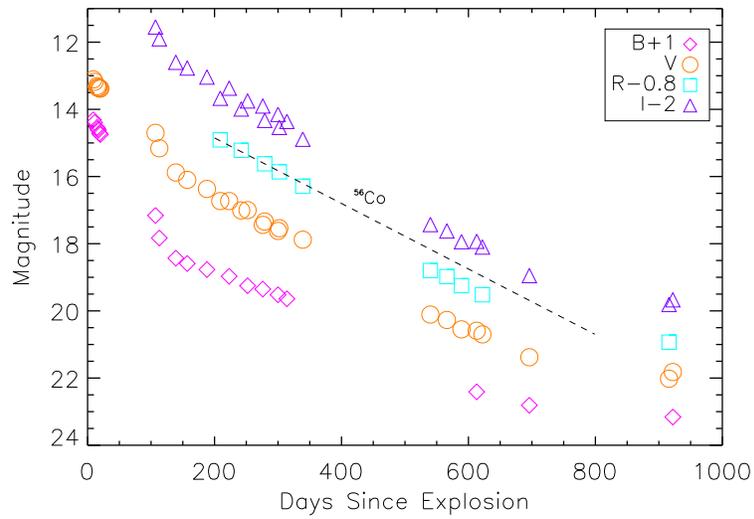}
  \caption{Optical BVRI lightcurves of SN 2007it. Photometry before day 100 was obtained from Carnegie Supernova Project.  The disagreement in values in $I$ between days 200 and 400 is likely due to bandpass differences between SMARTS and Gemini filters.  }
  \end{figure} 

\begin{figure}[h]
\centering
  \includegraphics [width= 3 in,angle=90] {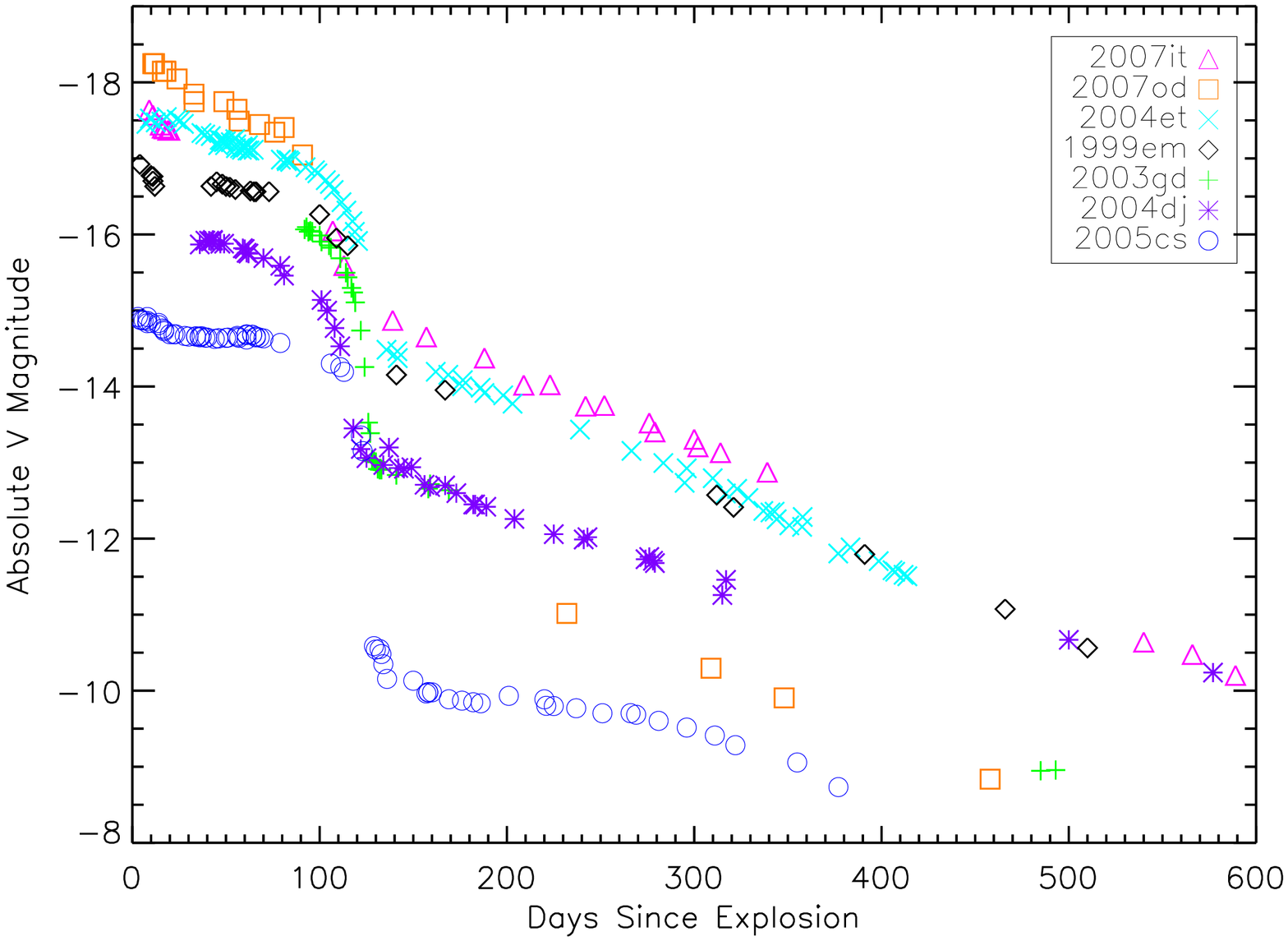}
  \caption{Absolute V light curves of a sample of Type IIP SNe. All SNe have been corrected for reddening and the data are from: SN 2007od \citep{2010ApJ...715..541A}, SN 2004et \citep{2009ApJ...704..306K,2007MNRAS.381..280M}, SN 1999em \citep{2003MNRAS.338..939E}, SN 2003gd \citep{2005MNRAS.359..906H}, SN 2004dj \citep{2006MNRAS.369.1780V,2006AJ....131.2245Z}, and SN 2005cs \citep{2009MNRAS.394.2266P}.}
  \end{figure} 

\begin{figure}[htp]
\centering
  \includegraphics [width= 3 in,angle=90] {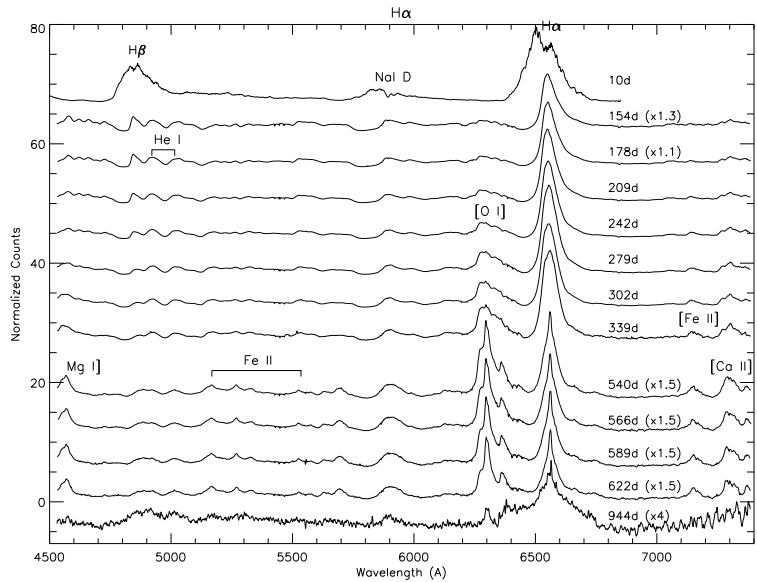}
  \caption{Spectral evolution of SN 2007it from day 10 to day 916.  All spectra other than day 10 are from Gemini/GMOS. Of interest is the brightness of the [O I] $\lambda\lambda$ 6300,6363 lines in relation to H$\alpha$ and [Ca II] lines. }
  \end{figure} 

\begin{figure}[htp]
\centering
  \includegraphics [width= 3.4 in,angle=90] {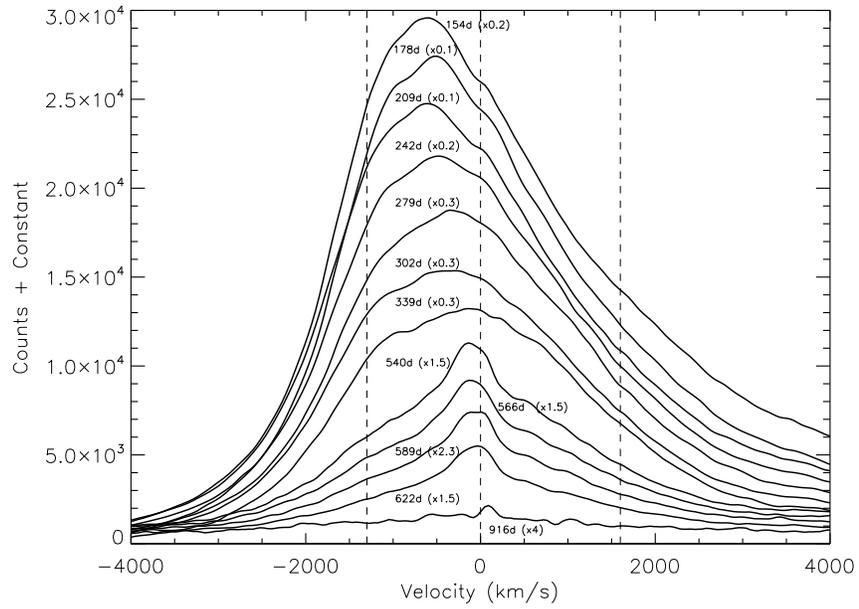}
  \caption{H$\alpha$ evolution from day 154 (top) to day 916 (bottom). Notice the emergence of an intermediate width component starting on day 540 which may indicate CSM interaction. }
  \end{figure} 

\begin{figure}[h]
\centering
\includegraphics [width=3 in,angle=90] {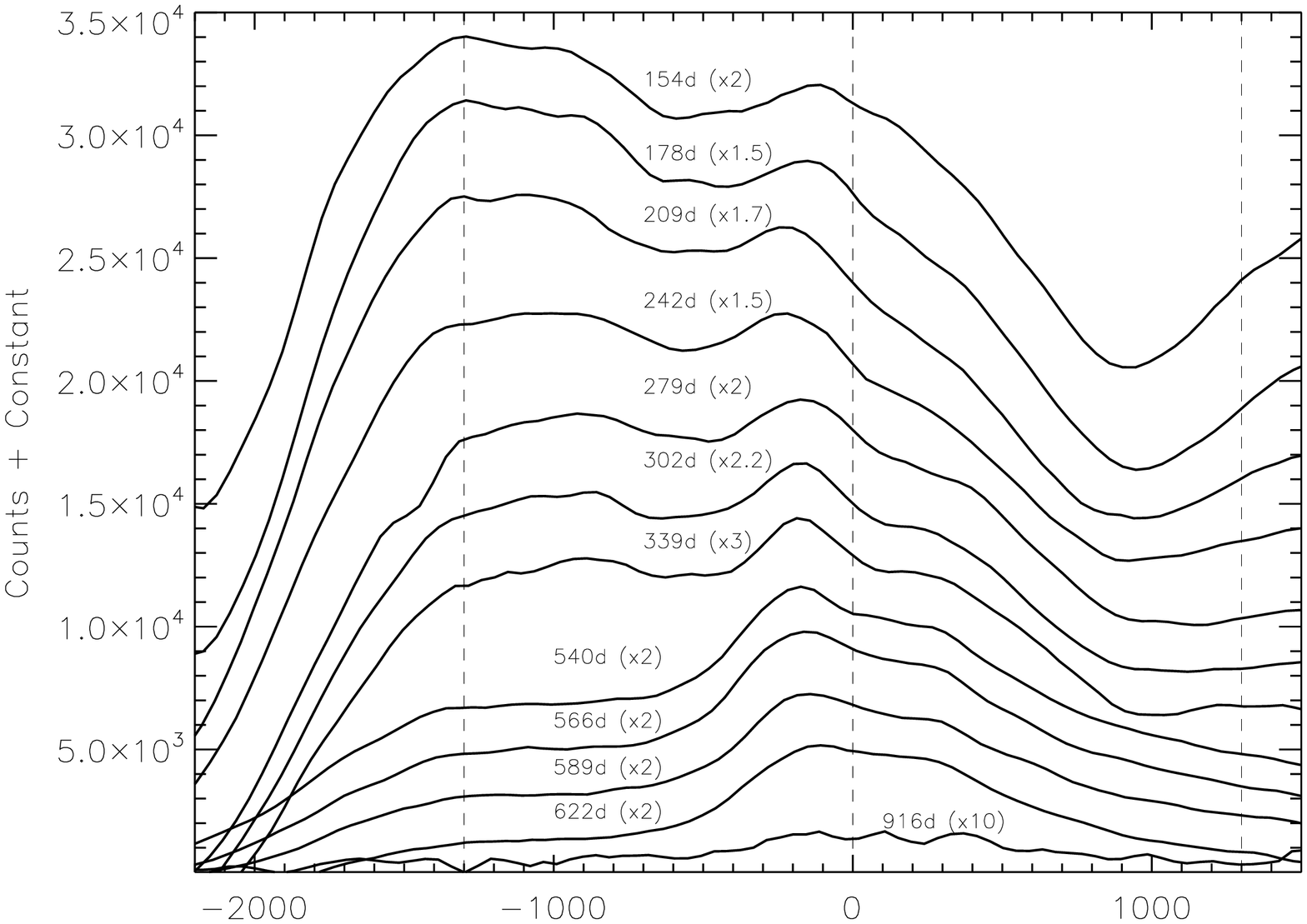}
\includegraphics[width=3 in, angle=90]{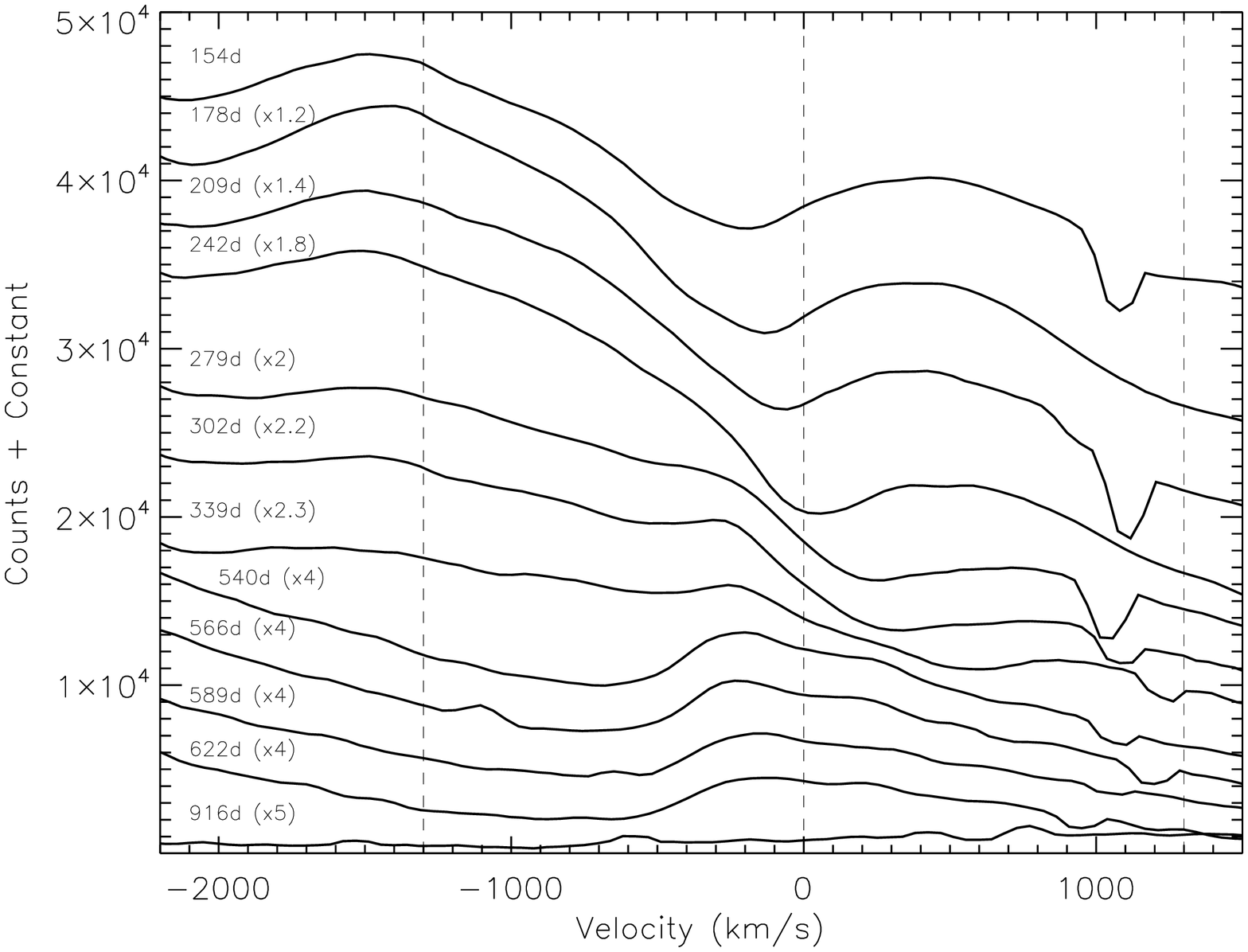}
\caption{ Evolution of [O I] 6300 \AA\ (top) and 6364 \AA\ (bottom) lines from day 154 to day 916. There is a separate blue-shifted component at -1300 km s$^{-1}$ from each [O I] line, which we believe is created by a torus of oxygen material surrounding the SN or a blob of O-rich ejecta on the forward side of the SN. }
\end{figure}

\begin{figure}[h]
\centering
\includegraphics [width=3 in,angle=90] {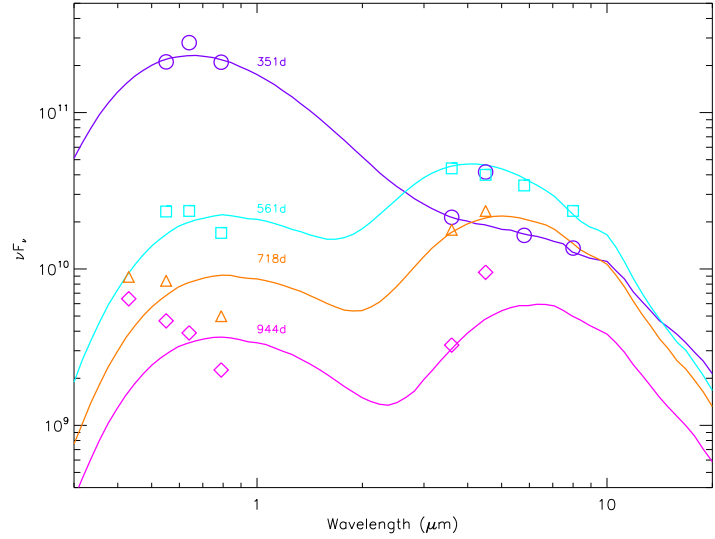}
\caption{MOCASSIN fits from our smooth shell modeling scenario (Table 6) for days 351 (circles), 561 (squares), 718 (triangles), and 944 (diamonds).  The optical points are from day 339, 566, 696, and 916/922 (Table 4), and have been corrected for foreground extinction.  Error bars are equal to or smaller than symbol size.  On days 718 and 944 a large portion of the optical luminosity is likely derived from the light echo, which due to its scattering nature will make the SN appear to be more blue \citep{2005MNRAS.357.1161P}.  The clumpy shell and torus modeling is consistent with the smooth shell fits, and are not shown.}
\end{figure}

\end{document}